# SLOW EXTRACTION BEAM COMMISSIONING FOR THE Mu2e EXPERIMENT AT Fermilab*

V. Nagaslaev†, G.Annala, J.Berlioz, G.Deinlein, B.Drendel, J.Morgan, A.Narayanan, J.StJohn, D.Still, D.VanderMeulen

Fermi National Accelerator Laboratory, Batavia, United States

*Abstract*

Following the successful completion of the Muon g-2 experiment run at Fermilab, the Muon Campus facility has been reconfigured from delivering 3 GeV muon beams to the g-2 storage ring to providing slow-extracted 8 GeV proton beam spills for the Mu2e experiment. The first full-scale commissioning run with slow extraction was conducted during the 2025 run, followed by the second run in early 2026. We present the results and current status of this commissioning campaign.

## BEAM REQUIREMENTS

The Mu2e experiment [1] requires a slowly extracted 8 kW proton beam delivered in short (<200 ns) pulses separated by ~1.6 μs intervals. To achieve this, beam is slowly extracted from the Delivery Ring (DR), where a single harmonic H = 4 RF bunch circulates with a revolution period of 1.695 μs. The machine cycle begins with two batches of 84 bunches at 53 MHz transferred from the Booster to the Recycler. After fast rebunching to 2.5 MHz in the Recycler, the resulting eight bunches are transferred bucket-to-bucket to the DR, one at a time.

The design parameters are summarized in the Table 1.

Table 1: Original Design Beam Parameters

|  |  |
|---|---|
| Beam kinetic energy | 8GeV |
| Beam intensity at injection | 1e12 |
| Cycle time | 1.4 sec |
| Spills per cycle | 8 |
| Spill length | 53ms |
| Beam power | 8kW |
| Spill Duty Factor | >60% |
| Extraction efficiency | >98% |

While beam intensity and time structure are dictated by experimental requirements, beam loss control is critical for machine operation. Due to radiation constraints, excessive losses may significantly limit the achievable beam power.

## DELIVERY RING AND SX

The Delivery Ring is a reconfigured version of the Debuncher ring from the former Fermilab Antiproton Source [2], adapted for the Muon Campus, which serves both the g-2 [3] and Mu2e experiments. Following completion of the g-2 program, a new beamline to the Mu2e experimental hall was constructed and a Slow Extraction (SX) system implemented. The original Antiproton Source optics has been preserved, with modifications to support slow extraction. Figure 1 shows the Delivery Ring lattice functions. The ring circumference is 505 m and consists of three arcs and three straight sections (SS). The lattice follows a regular FODO structure with maximum beta functions of 15 m in both planes. Dispersion reaches up to 2 m in the arcs and is suppressed in the straights using the "missing dipole" scheme

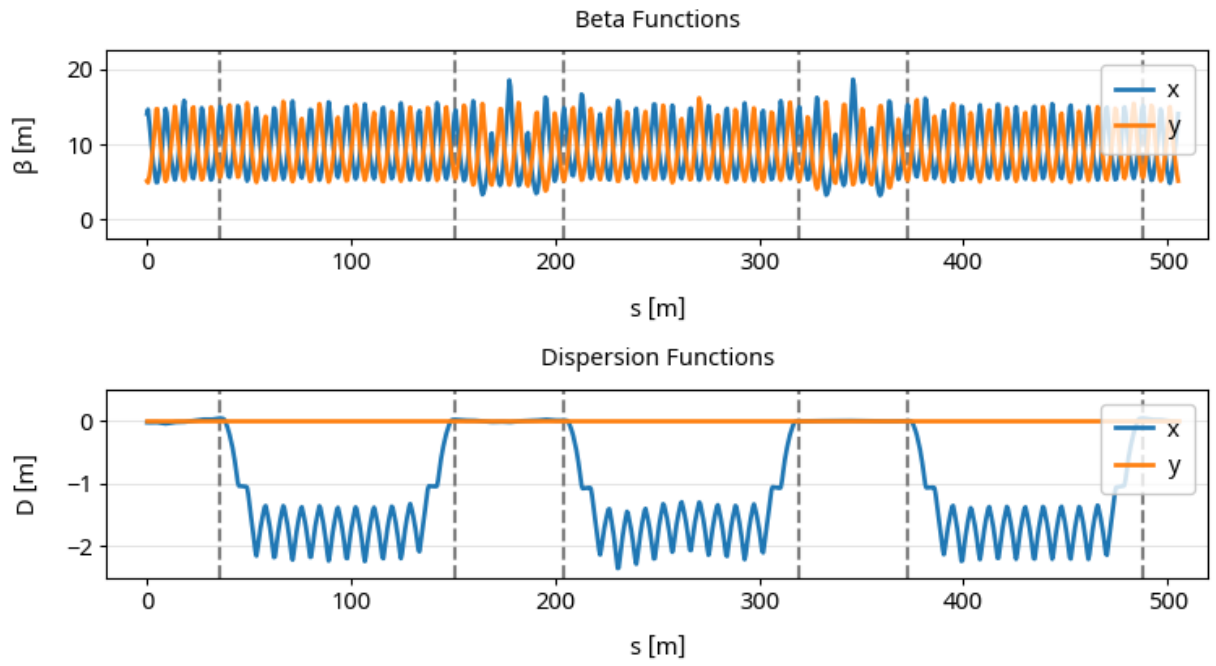

Figure 1. Lattice functions, optimized for SX

Figure 2 shows the main features of the new machine:

**Extraction system:** Injection and extraction are both located in SS30. While the injection scheme remains unchanged from g-2 operations, Mu2e employs two electrostatic septa (ESS1 and ESS2) instead of a single-pass kicker.

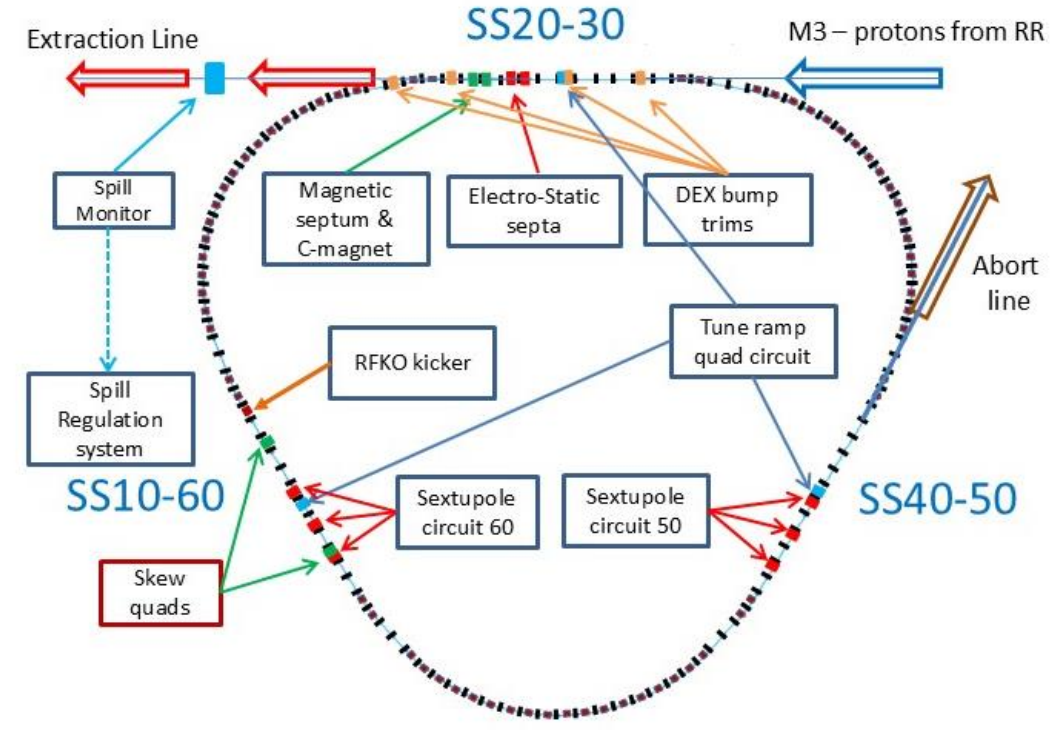

Figure 2. Delivery Ring configuration with SX elements.

The two ESS have 12 mm gap and operate at 86 kV. Together they provide a horizontal deflection of 2 mrad. The septum planes consist of 25 μm thick, 1 mm wide W/Re foils with 2.5 mm spacing. Downstream, two magnetic septa—a Lambertson magnet (ELAM) and a C-magnet (CMAG)—deflect the extracted beam vertically into the beamline. With an additional contribution from quadrupole Q205, the total vertical bend is 112 mrad.

* This work is supported by Fermi Forward Discovery Group, LLC, acting under Contract No. 89243024CSC000002.

†vnagasl@fnal.gov

**Two sextupole circuits,** SXCOS and SXSIN, are installed in SS10 and SS50. Each circuit includes three sextupoles positioned near focusing quadrupoles, with phase advances of ~60°. The circuit locations are chosen to produce nearly orthogonal contributions to the third-integer resonance term. Each circuit is powered by an independent fast-switching power supply, providing ~10 ms full ramp time.

**The tune ramping circuit** (QX) consists of three quadrupoles located in the middle of each straight section. Extraction is achieved by exciting the third-integer resonance at $Q_{res}$=29/3. The tune is ramped from 9.65 toward resonance using the QX circuit. To facilitate the fast machine reset back within 5ms, the QX fast switching supply has been augmented with the high voltage regulation module.

**Dynamic Orbit bump (DEX):** A dynamic orbit bump is built to provide fast extraction orbit corrections during the spill. It uses four dipole correctors in SS30 and can be configured to provide angle and/or position bumps.

**The Radio-Frequency Knock-Out system (RFKO)** has been built to assist extraction with fast beam heating. For this purpose, the 1.5m long strip-line kicker has been installed in the SS10. The kicker is driven by the two solid state TOMCO PA-s with total max power 1.4kW.

**The Spill Regulation System (SRS)** has been designed and built to control the spill process and regulate the spill performance in real time. The system has been implemented in the Altera Arria-10 SoC system, capable of delivering up to 10kHz decision rate for fast feedback regulation. Not all the design functionality has been implemented in the board at the time of first runs.

The block diagram of the SRS is shown in Figure 3. SRS is driving the QX ramp to manage the spill squeeze. The fast spill feedback corrections are superimposed on this ramp. Alternatively, fast regulation can be achieved by manipulating the extraction rate using the RFKO beam heating. The fast regulation is driven by the PID Controller.

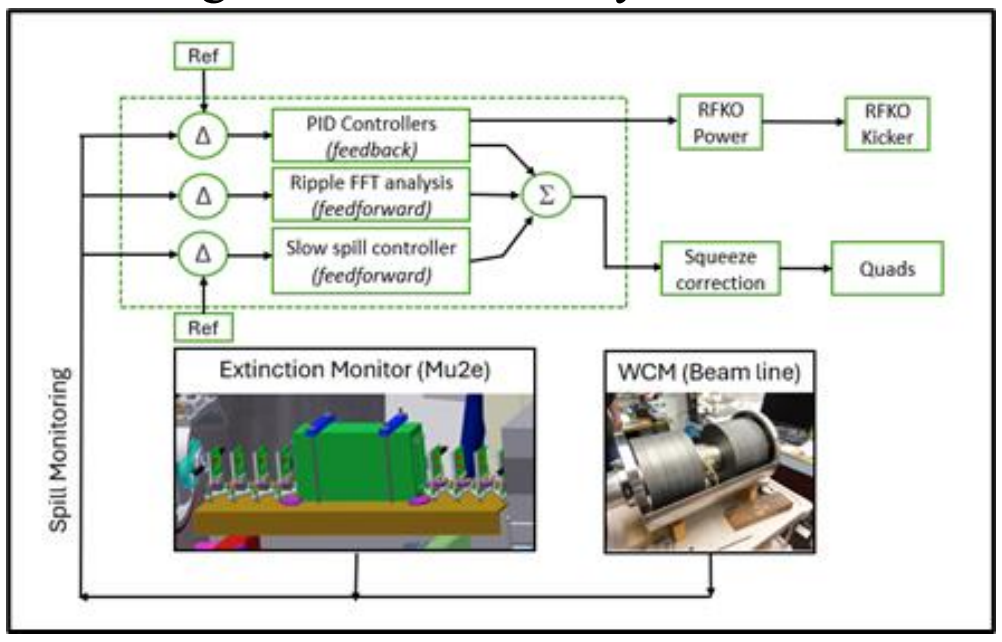

Figure 3. Block diagram of the Spill Regulation system.

Two other controllers in the SRS board provide the slower, feed-forward regulation. The Slow spill controller analyses the average spill profile over many spills and applies correction to the current QX ramp curve stored on the board. The ripple controller analyses the harmonic and phase content of the ripple structure and produces the feedforward corrections to the squeeze. It may also send the feedforward correction waveforms to the relevant power supplies.

## BEAM COMMISSIONING

### *Resonance control*

Optimization of the resonance term is achieved by aligning the separatrix orientation at ESS1. The phase space at the ESS1, reconstructed from the TBT BPM data is shown in Figure 4, in normalized coordinates.

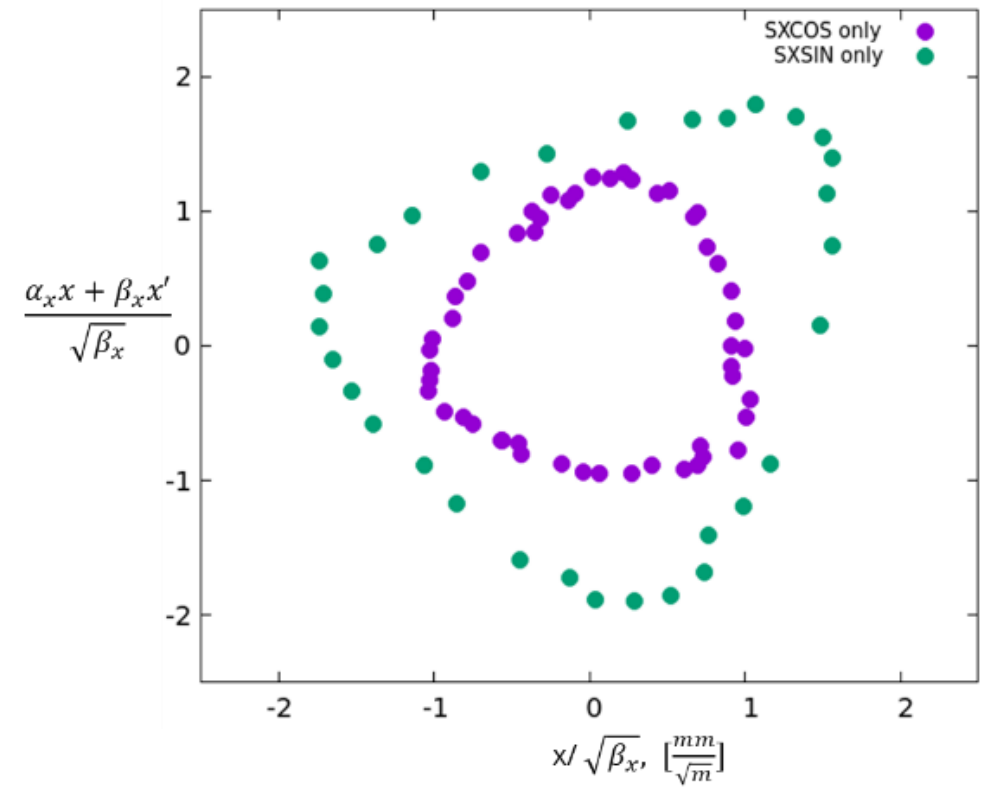

Figure 4. The phase space beam trajectories at the presence of SXCOS or SXSIN fields only.

Measurements indicate that the SXCOS sextupole configuration is already close to optimal. Maximum proton yield to the absorber can be achieved by addition of SXSIN component at 10% level. Because the yield improvement in this case is quite marginal, commissioning has so far relied primarily on SXCOS.

### *The Spill Regulation System*

In the absence of the Mu2e Extinction Monitor yet, the spill characteristics have been monitored using the Wall Current Monitor (WCM) in the extraction beam line M4. Figure 5 is showing a typical time structure of the WCM response (red trace).

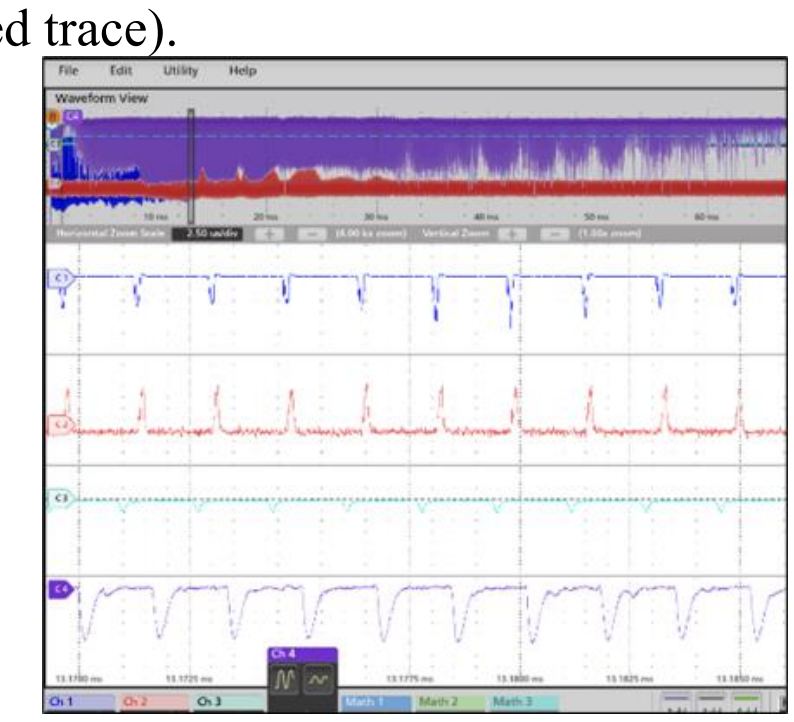

Figure 5. Scope screenshot with zoomed traces. The WCM trace is in red. Full zoom window width is ~16 microsec.

The pulse structure reflects the RF time structure of the circulating beam, with the width about 200nsec and separation of 1695nsec. For the purposes of spill regulation, the WCM waveform is digitally integrated in the FPGA into

** Fermilab style magnets and power supplies

currently supported bandwidth (here 10kHz, as shown in Figure 7).

The slow spill regulation controller at this time has been temporarily implemented externally, using piecewise ramp tuning algorithm.

Figure 6 shows evolution of the QX ramp from the initial "guess" logarithmic ramp curve (left) to the final optimized ramp on the right. Figure 7 shows improvement in the spill profile. The typical spill duty factor of the optimized squeeze is above 90%.

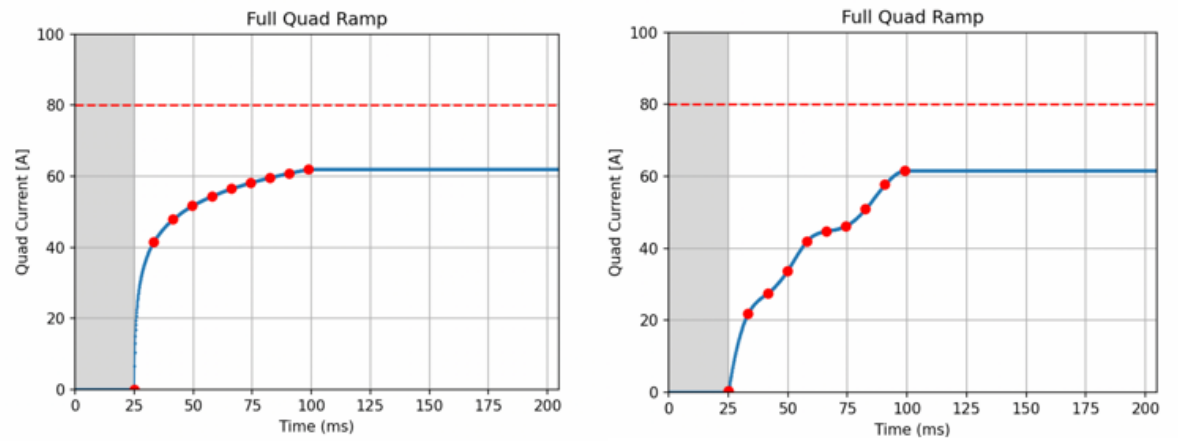

Figure 6. Spill flattening. Left: initial log-curve ramp; Right-optimized QX ramp.

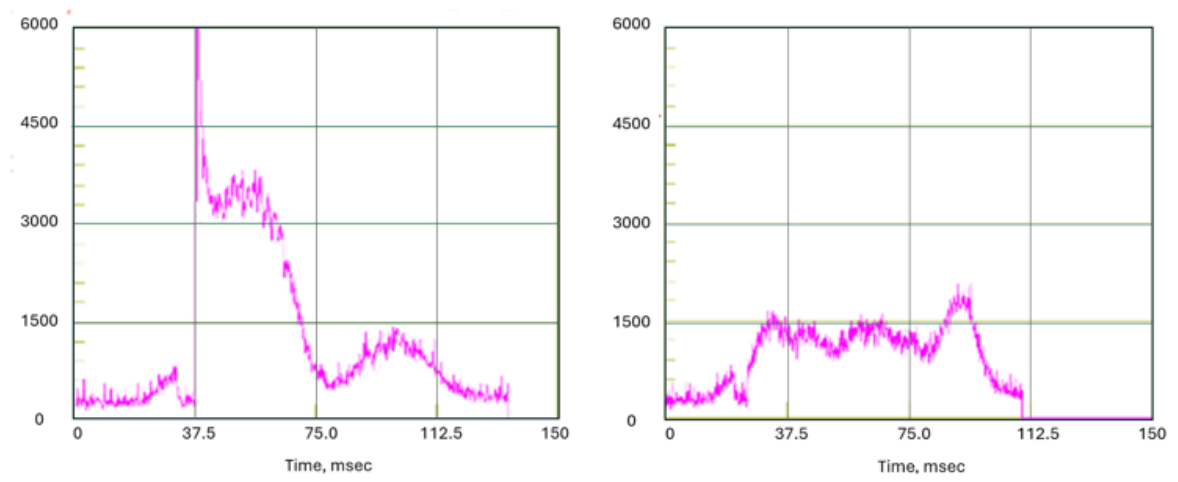

Figure 7. Spill flattening. Spill profiles at 10kHz. Left-initial curve; right-optimized profile.

### *Beam losses*

Normally, the beam losses in SX are dominated by the septum losses. In our case the septum losses are estimated to be about 2% and this can be further reduced. However, there still remain several sources of other beam losses in the machine. Currently, the beam losses limit the available beam power to target at about 5kW. One of the sources is the steering errors in the beam lines. Correction of these errors is currently impeded by the availability of the beam diagnostics and that will be addressed soon. The other sources are not well understood at this time and will be in the main focus of the following runs.

### *Multi-spill mode*

DR is in the process of obtaining the safety clearance for the 8kW operation. The current power limit is only 13W, determined by the legacy design of the Antiproton Source. This only allows running a few individual spills per minute, with intensity reduced by a factor of 5. However, as a proof of principle we also demonstrated a multi-spill mode, running 4 spills during a Main Injector cycle; the design mode of 8 spills can be achieved by extending this sequence.

### *Dynamic bump*

The DEX bump has been implemented to stabilize the extraction angle at the ESS1. Extraction angle drift during the spill is substantial and this correction is critically important for keeping the septum losses low. The dynamic bump was obtained by building a closed angle bump and scaling all the dipole currents in time according to the computed ramp curve. It was verified statically that the bump remains closed in the entire range of the ramp curve.

Figure 8 shows the water-fall display of the beam centroid measured in one of the beam line beam profile monitors. Horizonal axis shows the wire number (1mm pitch) and vertical axis shows time in msec. The upper trace shows beam position shifting during the spill as a result of the extraction angle drift. The trace below is stable due to the extraction angle compensation by the DEX bump.

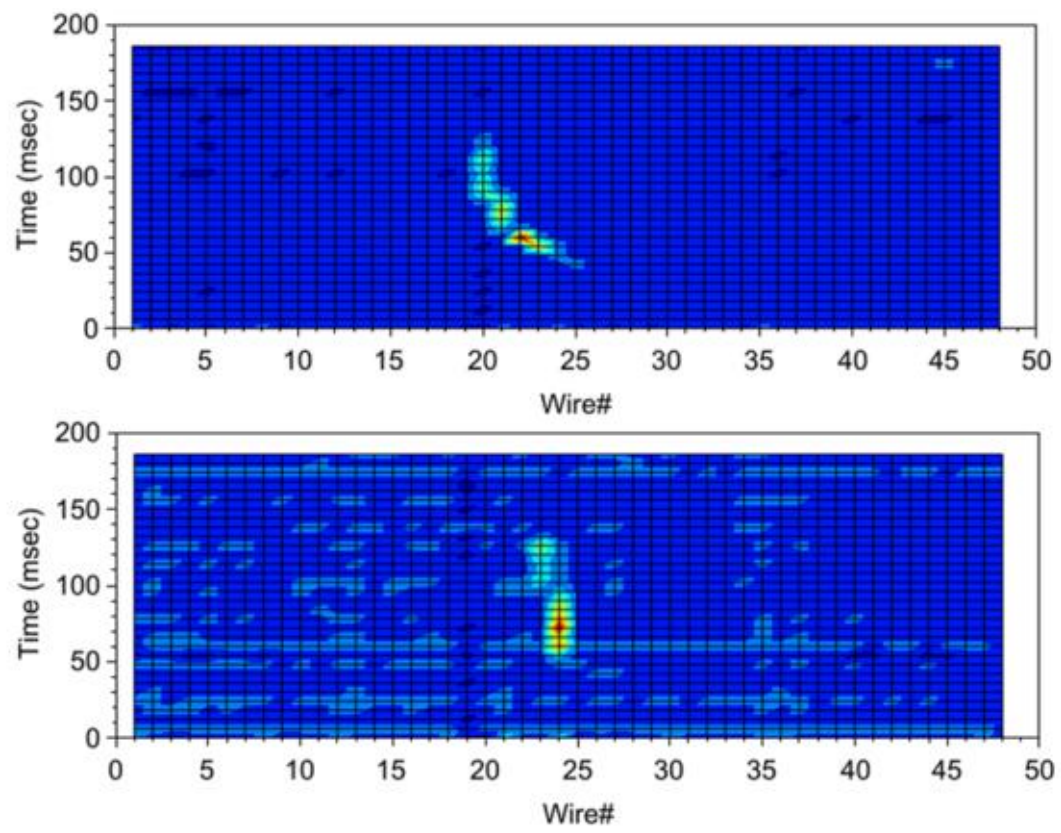

Figure 8. The beam centroid position in the beam line without (above) and with (below) the DEX bump.

## CONCLUSION

The Fermilab Delivery Ring has been reconfigured to deliver high intensity proton beams to the Mu2e Experiment using Slow Extraction. Commissioning of the machine and beam line transport has successfully begun. Slow Extraction has been established with beam delivered to the Diagnostic Absorber. ESS related losses are estimated to be better than 2% and will be further improved. The Spill Duty factor has been shown to reach over 90%. Multi-spill operation mode and efficient orbit stabilisation using DEX bump have been demonstrated. With that, there is a significant list of improvements that need to be addressed before the Mu2e experiment run, expected to begin in late 2027. The most significant items in this list are non-ESS beam losses, limiting available beam power to about 5kW, orbit management, fast timing control and spill regulation in the multi-bunch mode, etc.